# Identification and Mitigation of False Data Injection using Multi State Implicative Interdependency Model (MSIIM) for Smart Grid

Sohini Roy, Arunabha Sen
School of Computing, Informatics and Decision System Engineering
Arizona State University
Tempe-85281, Arizona, USA
Email: {sohini.roy, asen}@asu.edu

*Abstract*—Smart grid monitoring, automation and control will completely rely on PMU based sensor data soon. Accordingly, a high throughput, low latency Information and Communication Technology (ICT) infrastructure should be opted in this regard. Due to the low cost, low power profile, dynamic nature, improved accuracy and scalability, wireless sensor networks (WSNs) can be a good choice. Yet, the efficiency of a WSN depends a lot on the network design and the routing technique. In this paper a new design of the ICT network for smart grid using WSN is proposed. In order to understand the interactions between different entities, detect their operational levels, design the routing scheme and identify false data injection by particular ICT entities, a new model of interdependency called the Multi State Implicative Interdependency Model (MSIIM) is proposed in this paper, which is an updated version of the Modified Implicative Interdependency Model (MIIM) [1]. MSIIM considers the data dependency and operational accuracy of entities together with structural and functional dependencies between them. A multi-path secure routing technique is also proposed in this paper which relies on the MSIIM model for its functioning. Simulation results prove that MSIIM based False Data Injection (FDI) detection and mitigation works better and faster than existing methods.

## I. Introduction

Remote monitoring of the power grid by means of the Supervisory Control and Data Acquisition (SCADA) have been in practice for a long time. In order to improve the remote monitoring of the power network and fructify the concept of a smart grid, current and voltage sensors are further updated to develop more advanced devices like Phasor Measurement Units (PMUs) [2] which have a sensing module, a data processing module, a memory and a communication module. Output from the processing module of the PMU is sent to the communication module and finally to the network to be sent to the control centers (CCs). These PMUs sample current and voltage phases at the rate of 48 samples per cycle and send 30 samples per second. CCs use these PMU data to perform a number of analytical tasks like state estimation, to estimate voltage stability margin, to validate generator model, generate a contingency list [3] for the network and so on. This brings the need for a fast and secure communication network to carry the PMU data from the substations to the CCs with minimum delay and in a secure manner.

In order to achieve these goals, researchers are trying to find the best suited design of the ICT infrastructure for smart grid systems. The authors of [1] provide a realistic design of the ICT infrastructure for smart grid with inputs from a power utility in the U.S. Southwest, but it is a completely wired network with fiber optic cables using SONET-over-Ethernet and Ethernet-over-DWDM. Such a communication network is neither cost effective nor energy saving as every entity draws power from the grid to operate. Furthermore, addition of new ICT components and isolation of the same is also difficult in such a network.

Since smart sensors like PMUs are already gaining popularity in the remote monitoring of power grid, a sensor network based ICT network design [4] became an easy choice for the researchers. Therefore, in this paper, a WSN based ICT network design for smart grid system is presented. Yet, WSNs are vulnerable to cyber-attacks and the proposed network design should be accompanied with an efficient and secure routing scheme to serve the purpose of pervasive monitoring of the power network. In order to do that, a clear understanding of the interactions between the different entities of the smart grid system is very important. In [1], an interdependency model, named as the Modified Implicative Interdependency Model (MIIM) is proposed that very accurately depicts the interdependencies between the different types of entities in the smart grid and also takes into account the interactions between them to determine the operational state of each such entity. MIIM uses Boolean logic based equations called Inter-Dependency Relations (IDRs) to model the structural and functional dependencies between entities in a smart grid. Just by solving such IDRs, the operational level of an entity can be identified. However, by solving IDRs it cannot be predicted if the data received from an operational entity is correct or it has false data injected into it. CCs completely depend on the data carried to it by the communication system from the PMUs to make all the required analysis. However, such data dependency is not covered in the dependency model MIIM [1].

On the other hand, cyber-attacks like False Data Injection (FDI) is very common in the communication system of a smart grid. Cyber science researchers mainly focus on recovering the actual data [5] after a FDI attack takes place to perform state estimation. However, such processes are not only complex but also, they are time consuming. A better way of arresting FDI attack can be identifying the attack before it reaches the CCs and thereby detecting the source of the attack, so that the source can be completely avoided for further communications. In this paper, an updated version of the MIIM is proposed as Multi-State Implicative Interdependency Model (MSIIM) which together with the structural and functional dependencies also takes into account the data dependency between the ICT entities of the smart grid. MSIIM ensures operational accuracy of the entities in the ICT network. A novel multi-path data routing technique is also proposed in this paper which prevents FDI attacks and even if an attack takes place, that can be identified in the path to the CC and by solving the MSIIM IDRs, the source of the attack can be detected.

The rest of the paper is structured as follows. Section II gives a brief description of the WSN based ICT network design for a smart grid of IEEE 14-Bus system. Section III gives an overview of the dependency model MSIIM. Routing of PMU data from the substations to the CCs is described in detail in section IV. Section V gives the simulation results and the paper concludes in section VI.

## II. DESIGN OF WSN BASED ICT NETWORK FOR SMART GRID

A generic ICT network design for smart grid system is proposed in this section considering IEEE 14-Bus system as the power grid. This design can be applied on larger power grids. Fig.1. shows all the power and communication entities in a smart grid of IEEE 14-Bus.

Before deciding on the design of the ICT network, the power grid is divided into a number of substations as in [1]. In Fig.1., there are a total of 11 substations denoted by $S_i$ where i is the substation ID. Now, in order to efficiently monitor the power grid, the whole network area is first divided into a number of regions. In order to do that, the connectivity of each of the substations located at the borders of the network region is calculated. Starting from a substation $S_i$ with the maximum connectivity among the border substations, all other substations within a given distance of it, are marked as substations of a common monitoring region $R_x$. Then the next substation which is closest to $S_i$ but beyond the given distance and is not yet placed in a region, is selected and the same process is repeated. This process is continued till every substation is placed within a monitoring region. Fig 1. shows that after this process is completed, the smart grid network for IEEE 14-Bus system is divided into 4 regions.

The substation $S_i$ having the maximum connectivity with other substations is selected as the main CC for the smart grid and that having the second highest connectivity is selected as the backup CC. If the communication entities of the main CC fail, the backup CC can still continue the remote monitoring of the smart grid system.

The WSN is used in the proposed ICT design to send the huge volume of PMU data from the PMU containing substations to the CCs. PMUs are placed in some of the substations following the optimal PMU placement algorithm in [6]. In Fig.1, PMUs ($U_i$) are shown by green diamonds on the buses. Each substation has a substation gateway ($GW_i$) which is responsible for receiving PMU data and SCADA data from the PMUs and RTUs of that substation respectively. A region is provided with a Phasor Data Concentrator (PDC), only if it has a $U_i$ containing substation. In the smart grid of IEEE 14-Bus, PDCs are placed in region 1 and 3 only. Energy Harvesting Relay Nodes (EHRNs) denoted by $R_i$ in fig.1., are densely deployed over the network area, mainly near the PDCs, and the $U_i$ containing substations. $U_i$ data from the substations are carried to the PDC of that region by the EHRNs following a multipath routing technique discussed in section IV. There is a dedicated optical fiber channel from each PDC to each of the CC-gateways. The CCs are provided with servers which are connected to the CC-gateways via LAN. $U_i$ data from all over the smart grid is sent to the CC-servers as they are received by the CC-gateways. The utility operators rely on the CC-servers to perform different analysis including state estimation. The two CC-gateways are connected by a fiber optic cable.

## III. OVERVIEW OF MULTI STATE IMPLICATIVE INTERDEPENDENCY MODEL (MSIIM)

The smart grid system can be viewed as a joint power and ICT network. In both MIIM [1] and MSIIM, the smart grid can be represented as a set $J(E, F(E))$ where $E = P \cup C \cup CP$ denoting the set of all entities in the smart grid. P denotes the set of all power entities; C denotes the set of communication entities and CP denotes entities belonging to both the layers like PMUs. $F(E)$ represents the set of IDRs. All the power entities in both the models are denoted as P type entities where $P = \{P_1, P_2, \dots P_n\}$ and the communication entities are C type entities where $C = \{C_1, C_2, \dots C_m\}$. In MIIM, the set $F(E)$ only captures the structural and functional dependencies between the entities in the smart grid.

In MSIIM, together with structural, functional or operational dependencies, the data dependency and operational accuracy is also taken into account while formulating the IDRs in set $F(E)$. In MIIM, any entity can take a value of 0, 1 and 2; indicating no operation, reduced operation and full operation respectively. The novelty of MIIM lies in the fact that it considers a reduced operational level of the smart grid entities which is a very common feature of them. However, even if a communication entity is fully operational, it may not deliver correct data to the entity in the next hop due to node compromise by attacker. Also, the CP type entities like $U_i$s can send wrong readings to the CCs. This might lead to a wrong analysis of the system state by the CCs and thereby compel the operator to take wrong decision. In MSIIM, the P type entities can take the same 3 states as described in MIIM, but the state values are 3 for full operation, 2 for reduced operation and 0 for no operation. The C and CP type entities can take four different states in MSIIM as the reduced operation is further categorized into two types, namely– operation reduced by false data and operation reduced by interdependent non-operational entities. Therefore, the four different states of C and CP type entities in MSIIM are denoted by– 0 indicating no-operation, 1 indicating reduced operation by false data, 2 indicating reduced operation by interdependent non-operational entities and 3 indicating full operation.

In MSIIM the same three new Boolean operators are used as MIIM [1], except the inputs and outputs can range from 0 to 3 in place of 0 to 2. The truth table for the 3 operators in MSIIM are given in Table I.

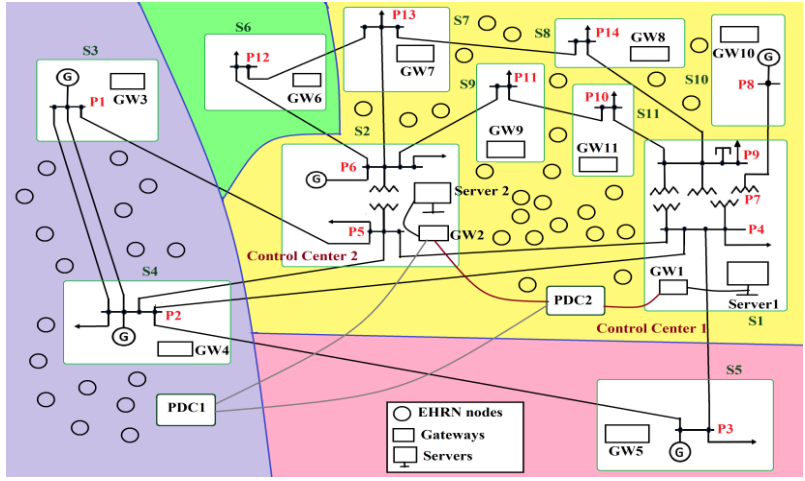

Fig. 1. Power and Communication Entities in a smart grid system of IEEE 14-Bus

In order to explain the difference between the two dependency models, a dependency between C type and P type entities is considered as follows. $C_i$, a C type entity, be operational if (i) $C_j$ which is another communication entity and $P_a$ which is a P type, are operational, or (ii) $C_k$ which is another C type entity and $P_b$ which is another P type entity are operational, and (iii) $C_l$ which is an entity in communication layer is operational. Now, let us assume that even if an entity in condition (i) or (ii) fails, $C_i$ will still work full operability, but if (iii) is not satisfied then $C_i$ will operate at a reduced level; this relation can be expressed using MIIM an MSIIM IDRs as: $C_i \leftarrow \left((C_j \circ P_a) \bullet (C_k \circ P_b)\right) \odot C_l$.

TABLE I. TRUTH TABLES FOR MSIIM OPERATORS

| Input 1 | Input 2 | Max_OR | Min_AND | New_XOR |
|---|---|---|---|---|
| 3 | 3 | 3 | 3 | 3 |
| 3 | 2 | 3 | 2 | 2 |
| 3 | 1 | 3 | 1 | 1 |
| 3 | 0 | 3 | 0 | 2 |
| 2 | 2 | 2 | 2 | 2 |
| 2 | 1 | 2 | 1 | 1 |
| 2 | 0 | 2 | 0 | 2 |
| 1 | 1 | 1 | 1 | 1 |
| 1 | 0 | 1 | 0 | 1 |
| 0 | 0 | 0 | 0 | 0 |

TABLE II. EVALUATION OF IDRs TO OBTAIN STATE VALUES

| | MSIIM | MIIM |
|---|---|---|
| STEP 1 | $C_l \rightarrow 1$ | $C_l \rightarrow 2$ |
| STEP 2 | $C_i \leftarrow (((3 \circ 3) \bullet (3 \circ 3)) \odot 1)$ | $C_i \leftarrow (((2 \circ 2) \bullet (2 \circ 2)) \odot 2)$ |
| STEP 3 | $C_i \leftarrow ((3 \bullet 3) \odot 1)$ | $C_i \leftarrow ((2 \bullet 2) \odot 2)$ |
| STEP 4 | $C_i \leftarrow (3 \odot 1)$ | $C_i \leftarrow (2 \odot 2)$ |
| STEP 5 | $C_i \leftarrow 1$ | $C_i \leftarrow 2$ |

In the first step of the Table II itself, the false data injection by $C_l$ is not captured by MIIM as the entity is fully operational otherwise. After evaluating the IDR also, this drawback of $C_l$ is not reflected in the state value of $C_i$ which depends on the data from $C_l$. However, in MSIIM, all the entities getting data from $C_l$ will be having a state value of 1 and this will help the destination node or a PDC to trace back through the path of entities having a state value 1 and identify the entity injecting false data into the system. Yet, the challenge lies in detecting the false data injection by a particular node or ICT channel. In this paper, it is assumed that, a node injecting false data into the system cannot identify that but, all nodes receiving data from that compromised node can identify the false data injection using the technique described in section IV and thereby, they update the state value of the node in the previous hop to 1. Unlike MIIM IDRs which are evaluated by the CCs based on data received from the RTUs and PMUs, each communication entity of the smart grid in MSIIM can evaluate their own IDR and send the state value to the CC. Moreover, IDRs are re-generated for only the communication entities which are selected for sending data to the CC in that round of data routing. Therefore, the IDR for a C type entity consists of other C type entities which are also selected for that round. Only the IDRs of P type entities are evaluated by the CCs. A distributed state table is maintained in MSIIM where each entity has the state values of all its connected entities. Each time data is received from some of the C types entities connected to it and data successfully sent to the next hop, the state table is updated. In case of a false data injection, the compromised node will have a state value of 2 or 3 and all its neighboring C type entities getting data from it will have its state value as 1. This will help the PDC to identify the compromised node. Section IV illustrates this with the help of fig.3.

IV. ROUTING OF PMU DATA FROM SUBSTATIONS TO CONTROL CENTERS

A. Assumptions

- PMUs, CC-gateways and CC-servers are trusted.
- FDI attack can take place at any point after the data is sent by the PMUs and before it is received by the CC-gateways.
- An MSIIM state table is maintained by each communication entity in the smart grid and whenever,

an entity gets involved in the routing process, it recalculates its own state value based on the state values of entities connected to it and the MSIIM IDRs.

### B. Secure Routing Technique

The MSIIM based secure routing technique for a smart grid system is divided into three modules as follows:

*1) Module 1: Data forwarding to substation gateways by PMUs*

The PMUs generate 48 time stamped samples of data per processing unit clock cycle from the analog signals received from the current and voltage sensors. The samples are sent to the $GW_i$ in which the PMU is placed via a dedicated wireless communication channel. In this work, PMUs are considered trusted. Data from the PMUs of a substation are stored in separate data queues by the $GW_i$. The MSIIM IDR for a substation gateway is given as: $GW_i \leftarrow [(PMU_1 \circledcirc PMU_2 \circledcirc ... PMU_n) \circ (R_1 \circledcirc R_2 ... R_m)]$ assuming there are n number of PMUs in the substation and m number of relay nodes at one-hop distance from the $GW_i$. If the $GW_i$ receives data from all the PMUs of that substation then it works at state 3. If any one of the PMUs don't operate, then its state changes to 2. Similarly, it should remain connected to all the relay nodes at one hop distance to get a state value 3. After calculating its own state value, the $GW_i$ starts Module 2 of the secure routing technique.

*2) Module 2: Data forwarding by substation gateways to PDCs*

The steps of module 2 is shown in the flowchart of fig.2. In this module, each gateway $GW_i$ discovers all possible paths from itself to the PDC of that region, consisting of nodes having a state value $\geq 1$. $GW_i$ calculates the trust values of each path by forwarding a number of test messages through them. Trust value of the i<sup>th</sup> path is calculated as:

$$TV_i = \sum_{j=1}^{nodes\ in\ path\ i} \frac{test\ messages\ delivered\ by\ node\ j}{test\ messages\ sent\ to\ node\ j} \quad (1)$$

$GW_i$ then selects four paths with the highest $TV_i$. Pair-wise keys are generated and exchanged between adjacent nodes in each path and also the MSIIM IDR of each node in a path is also generated as shown in fig.2. $GW_i$ now generates a secret key $SK_i$ and shares it with the PDC of that region :

- Four random binary numbers of equal size– $k_1, k_2, k_3$ and $k_4$ are generated by $GW_i$.
- A secret key ($SK_i$) is now generated by $GW_i$ by XORing the four binary numbers:
$$SK_i = k_1\ XOR\ k_2\ XOR\ k_3\ XOR\ k_4 \quad (2)$$
- Now, each such binary number $k_i$ is XORed with a binary number with equal number of 1s to generate a key fragment in the following way:
    - Let, $k_i$ has a size 4: $k_i = B_1 B_2 B_3 B_4$ where $B_i$ is a binary digit.
    - Key Fragment ($KF_i$) is generated as: $KF_i = k_i\ XOR\ 1111$

- Now, each of 4 key fragments is encrypted using RC5 symmetric cypher and the pairwise key that the gateway shares with each of the next of nodes in the four most trusted paths. A Hashed Message Authentication Code (HMAC) [2] is generated over the total message using the same pairwise key.

- Each node in the trusted path matches the HMAC using the pairwise key it shares with its previous node in the path and if match is found, a new HMAC is generated over the encrypted key fragment using the pairwise key of the next node and this process is repeated till the encrypted key fragment reaches the PDC.

- The PDC receives the four key fragments via the four trusted paths and generates the secret key in the following way:
$$SK_i = KF_1\ XOR\ KF_2\ XOR\ KF_3\ XOR\ KF_4 \quad (3)$$

The Key Fragment sent via i<sup>th</sup> path is stored by each node in that path. Any attacker cannot get hold of the secret key even if it gets a Key Fragment.

The $GW_i$ now uses this secret key to encrypt each part of the data from PMU using RC5 symmetric cypher and it generates a HMAC using the same secret key over the encrypted data. $GW_i$ now selects one of the four most trusted path to forward the encrypted data. A nested HMAC is generated over the whole data using the $KF_i$ sent through that path and another nested HMAC is generated using the pairwise key it shared with the next hop nodes in the trusted paths. The encrypted and MACed data is forwarded to the next hop in one of the selected trusted paths.

The format of the data fragment sent via the i<sup>th</sup> path is:

$$GW_i \rightarrow R_i: EDF || HMAC\left(PK_i; HMAC(KF_i; HMAC(SK_i; EDF))\right)$$

where EDF is the encrypted data fragment. This process is repeated for all four parts of the 48 samples from a PMU and they are forwarded to the PDC using the four most trusted paths. The same paths are used by the $GW_i$ to forward data to PDC until an FDI attack is detected in one of the paths. The NHMAC using $PK_i$ helps in identifying FDI attacks by a compromised communication channel between two nodes; the NHMAC using $KF_i$ helps in identifying the FDI attack by any compromised node and any unobservable attack can be finally detected by the PDC using the NHMAC over the encrypted data generated using the secret key $SK_i$.

Each relay node $R_i$ receiving a data fragment from the previous node in the i<sup>th</sup> trusted path, first matches the HMAC generated using its own pairwise key with the previous hop node. If match is not found, it marks its own state value as 1. Now, if the HMAC generated using the key fragment $KF_i$ don't match, it also marks state value of the previous node in the path as 1. It stops forwarding the data fragment and sends its own state value to the next node. The next node calculates its own state value using MSIIM IDR and forwards the same to its next hop, and finally it is forwarded to the PDC. The PDC now uses the state values of the nodes to trace the node injecting false data [Fig.3.].

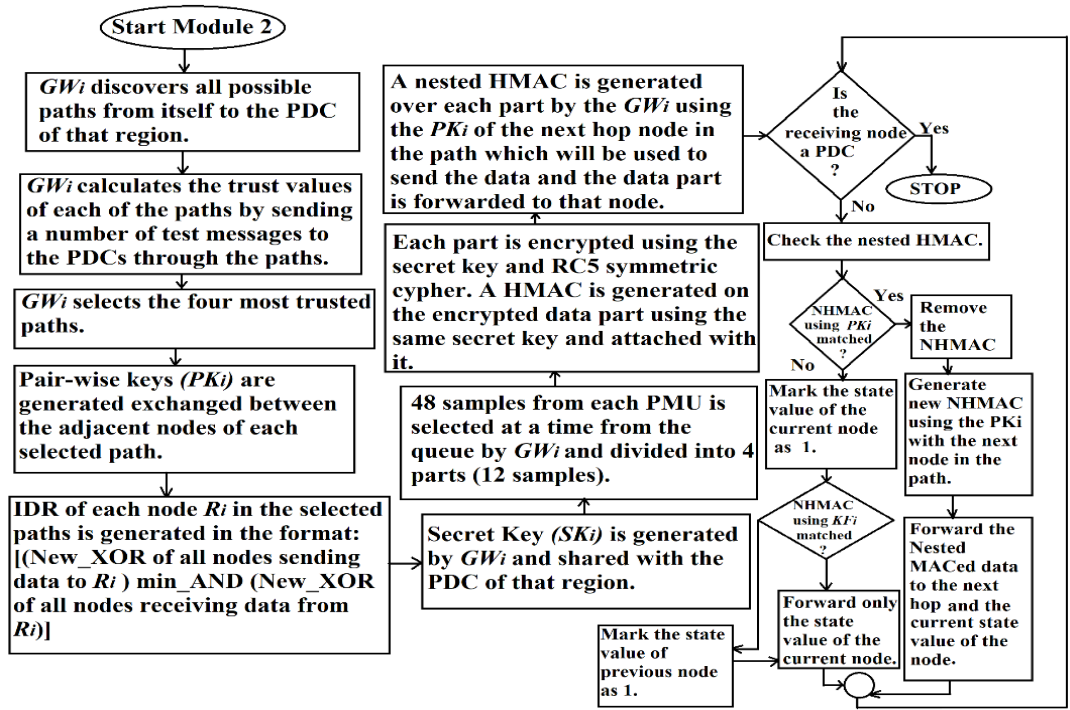

Fig. 2. Flowchart describing Module 2 of secure routing scheme

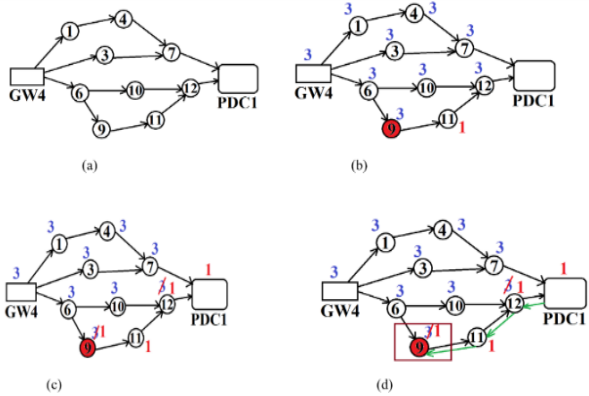

Fig. 3. Identification of the false data injecting node by PDC

In fig.3. (a), the trusted paths are: $\{R_1 \to R_4 \to R_7\}$, $\{R_3 \to R_7\}$, $\{R_6 \to R_{10} \to R_{12}\}$ and $\{R_6 \to R_9 \to R_{11} \to R_{12}\}$. In fig.3. (b), relay node $R_9$ is compromised and it is injecting false data into the system, which is first noticed by $R_{11}$ and it only forwards the state value to the next node $R_{12}$ and it also calculates its own state value in fig.3. (c). $R_{11}$ also changes the state value of $R_9$ to 1, in its own state table. The compromised $R_9$ will still maintain a state value greater than 1 in its own state table. PDC1 checks the state value of $R_{12}$ and it is 1 in both its own table and the state table of $R_{12}$. PDC1 now checks to the next-hop nodes connected to $R_{12}$. $R_{10}$ has a state value 3 and it is same in the state table of $R_{12}$ as well. There that path is not checked any more. PDC1 finds the state value of $R_{11}$ and $R_9$ as 1. When the state table of $R_9$ is checked, PDC1 finds that it has a state value greater than 1. Now $R_9$ is identified as the node injecting false data and the PDC immediately sends the ID of $R_9$ to both the CCs, so that $R_9$ is removed from the list of nodes in the ICT network of the smart grid. The PDC also sends the ID of $R_9$ and a NAK to $GW_i$. path with highest trust value from its stored list of discovered paths to resend that data fragment to PDC. In case a FDI attack is not detected by the relay nodes of a path but the NHMAC using the $SK_i$ generated by the PDC does not match with the attached NHMAC, then the whole path is marked as unsafe.

*3) Module 3: Data forwarding by PDCs to the CC-gateways*

Each PDC after receiving data from all the four paths decrypt them using the secret key. PDC aggregates the data from all the PMUs. Each of the CC-gateways use Elliptic Curve Diffie-Helman (ECDH) [2] key exchange scheme to establish a shared secret key with the PDC. The $GW_i$ uses this secret key to encrypt the aggregated data and it also generates a HMAC over the encrypted data using the same shared secret key and attach it with the aggregated. This data is now sent to the CC-gateways through the dedicated fiber optic cables.

V. SIMULATION RESULTS

In order to analyze the performance of the proposed secure routing technique that relies on the MSIIM model, a smart grid system of IEEE 118-Bus is considered. The network simulation area has 8 regions, 5 PDCs, 200 EHRNs and 107 substations. $S_{61}$ with buses $P_{68}$, $P_{69}$ and $P_{116}$ is selected as the main CC and $S_{16}$ having buses $P_{17}$ and $P_{30}$ is selected as the backup CC.

TABLE III. SIMULATION PLATFORM AND PARAMETER LIST

| Parameter | Description |
|---|---|
| Operating system | Fedora |
| Simulator | NS2.29 |
| Compiler | TCL |
| Rechargeable battery for EHSNs | NiMH |
| Communication standard | Zigbee |
| Battery capacity of EHSNs | 2000(mAh) |
| Initial battery power for all nodes | 200 (mAh) |

### A. Percentage of EHRN compromised vs. Communication delay

In fig.4, the communication delay vs. the percentage of node compromise is shown. When a certain number of malicious nodes are detected in the network region and the $GW_is$ have to rediscover new set of paths to the PDC, the communication delay increases. Again, when a number of paths are discovered and stored, increase in the number of malicious nodes only result in discarding of compromised paths and selection of new paths from the stored list, resulting in a slight drop of the communication delay. It is also observed that even with 25% of node compromise, the communication delay is as low as 30.4 seconds.

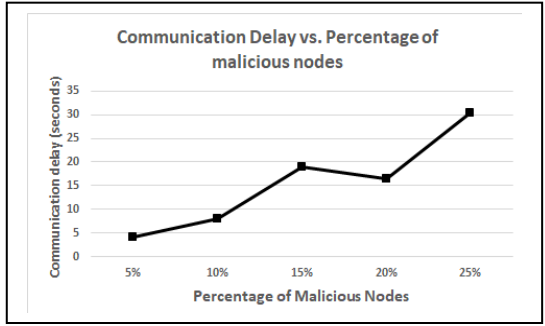

Fig. 4. Percentage of node compromise vs. Communication delay

### B. Number of EHRN compromised vs. Percentage of packets dropped

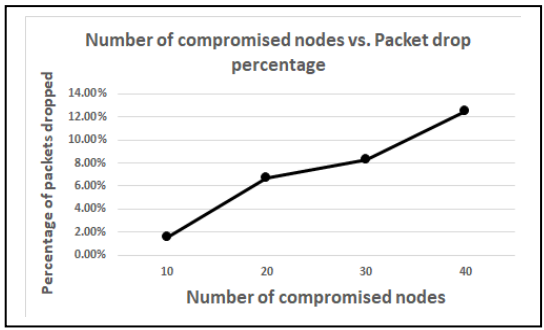

Fig. 5. Number of compromised nodes vs. Percentage of packets dropped

Fig.5. shows the number of compromised nodes in the network vs. the percentage of packets dropped. It is observed that even 40 EHRNs are compromised, the packet drop percentage is only 12.5%. Each gateway selects a portion of the total number of nodes in the network in the four most trusted paths to PDCs and as a result most of the other nodes remain idle. If such idle nodes are compromised, the communication of PMU data to the CCs will not be harmed and they will not be selected again. However, if an already selected node is compromised, it will result in packet drop in the next hop only. Then the false data is not forwarded further, and the faulty node is removed. Therefore, the packet drop is not very high in the proposed routing technique.

### C. Number of fabricated packets vs. Average energy consumed

In fig.6, the average energy consumed by each active node vs. the number of fabricated packets is shown. The average energy consumed by each active EHRN when FDI attack takes place, increases as the number of fabricated packets increase in the network. However, the consumption of energy is very nominal for resending the data packets that are fabricated before. When there are 45 fabricated data packets in the network, the average energy consumed by each node is 0.104 mAh.

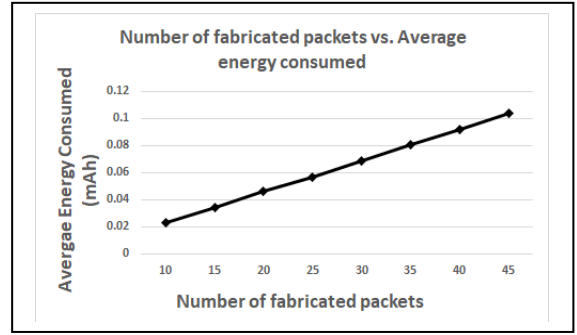

Fig. 6. Number of fabricated packets vs. Average energy consumed

### VI. CONCLUSION

The region based remote monitoring adopted in the proposed routing technique not only involves lesser number of relay nodes in forwarding the PMU data to the PDCs but also it helps in easy identification of a compromised node in a particular area of the smart grid. The MSIIM model proposed in this paper helps in accurate estimation of the operational levels of the entities in the smart grid. Forwarding of data packets to an ICT entity with lower operational level than required will result in increased packet drop and resending of packets. This model will also help the operators to examine a cyber-attack in the smart grid system before it actually takes place and know about the consequences. In the proposed work PMUs are considered to be trusted. Identifying false data injection by PMUs using the MSIIM model and thereby taking necessary actions can be a scope of future work.

### REFERENCES

[1] S. Roy, H. Chandrasekaran, A. Pal and A. Sen, "A New Model to Analyze Power and Communication System Intra-and-Inter Dependencies", *2020 IEEE Conf. on Tech. for Sustainability (SUSTECH 2020)*, Santa Ana, Apr. 2020, pp.181-188.

[2] S. Roy, "SSGMT: A Secure Smart Grid Monitoring Technique", in: Rashid A., Popov P. (eds) Critical Information Infrastructures Security. CRITIS 2020. Lecture Notes in Computer Science, vol 12332. Springer, Cham., 2020.

[3] S. Roy and A. Sen, "Identification of the K-most Vulnerable Entities in a Smart Grid System," *2020 3rd International Conference on Advanced Communication Technologies and Networking (CommNet)*, Marrakech, Morocco, 2020, pp. 1-6.

[4] A. Bashian, M. Assili, A. Anvari-Moghaddam, and J. P. S. Catalão, "Optimal Design of a Wide Area Measurement System Using Hybrid Wireless Sensors and Phasor Measurement Units," *Electronics*, vol. 8, no. 10, p. 1085, Sep. 2019.

[5] J. Liang, L. Sankar and O. Kosut, "Vulnerability Analysis and Consequences of False Data Injection Attack on Power System State Estimation," in *IEEE Transactions on Power Systems*, vol. 31, no. 5, pp. 3864-3872, Sept. 2016.

[6] A. Pal, A. K. S. Vullikanti and S. S. Ravi, "A PMU Placement Scheme Considering Realistic Costs and Modern Trends in Relaying," in *IEEE Transactions on Power Systems*, vol. 32, no. 1, pp. 552-561, Jan. 2017.